\documentclass[fleqn,figuresright]{2023SCGE}
\usepackage{hyperref}

\usepackage{multirow}
\usepackage{supertabular}
\usepackage{amsmath} 
\usepackage{amssymb}
\usepackage{makecell}
\usepackage{soul}
\usepackage{ulem}
\usepackage{longtable}
\usepackage{threeparttablex}

\newcommand{\Ep}{$E_{\rm \,peak}$}

\newcommand{\prl}{Phys. Rev. Lett.}
\newcommand{\apjl}{Astrophysical Journal Letters}
\newcommand{\apj}{Astrophysical Journal}
\newcommand{\apjs}{The Astrophysical Journal Supplement Series}
\newcommand{\mnras}{Monthly Notices of the Royal Astronomical Society}

\begin{document}

\ensubject{subject}

\ArticleType{Article}
\SpecialTopic{SPECIAL TOPIC: }
\Year{*}
\Month{*}
\Vol{*}
\No{*}
\DOI{??}
\ArtNo{000000}
\ReceiveDate{****}
\AcceptDate{****}

\title{The GECAM Ground Search System for Gamma-ray Transients}


\author[1,2]{Ce Cai}{caice@hebtu.edu.cn}%
\author[2,3]{Yan-Qiu Zhang}{}%
\author[2]{Shao-Lin Xiong}{{xiongsl@ihep.ac.cn}}%
\author[2]{Ping Wang}{{pwang@ihep.ac.cn}}%
\author[2]{Jian-Hui Li}{}%
\author[2]{Xiao-Bo Li}{}%
\author[2]{\\Cheng-Kui Li}{}%
\author[2]{Yue Huang}{}%
\author[2]{Shi-Jie Zheng}{}%
\author[2]{Li-Ming Song}{}%
\author[4,5]{Shuo Xiao}{}%
\author[2,6]{Qi-Bin Yi}{}%
\author[7]{Yi Zhao}{}%
\author[2,8]{\\Sheng-Lun Xie}{}%
\author[2]{Rui Qiao}{}%
\author[2]{Yan-Qi Du}{}%
\author[2]{Zhi-Wei Guo}{}%
\author[2,3]{Wang-Chen Xue}{}%
\author[2,3]{Chao Zheng}{}%
\author[2,3]{Jia-Cong Liu}{}%
\author[2,3]{\\Chen-Wei Wang}{}%
\author[2,3]{Wen-Jun Tan}{}%
\author[2,3]{Yue Wang}{}%
\author[2,3]{Jin-Peng Zhang}{}%
\author[2]{Chao-Yang Li}{}%
\author[2]{Guo-Ying Zhao}{}%
\author[2]{\\Xiao-Yun Zhao}{}%
\author[2]{Xiao-Lu Zhang}{}%
\author[2]{Zhen Zhang}{}%
\author[2]{Wen-Xi Peng}{}%
\author[2]{Xiang Ma}{}
\author[2]{Jing-Yan Shi}{}%
\author[2]{\\Dong-Ya Guo }{}%
\author[2]{Jin Wang}{}%
\author[2]{Xin-Qiao Li}{}%
\author[2]{Xiang-Yang Wen}{}%
\author[2]{Zheng-Hua An}{}%
\author[2]{Fan Zhang}{}%

\AuthorMark{Ce Cai}

\normalem
\AuthorCitation{C. C., Y.Q. Zhang, S.L. Xiong, P. Wang, et al.}

\address[1]{College of Physics and Hebei Key Laboratory of Photophysics Research and Application, Hebei Normal University, Shijiazhuang, Hebei 050024, China}
\address[2]{Key Laboratory of Particle Astrophysics, Institute of High Energy Physics, Chinese Academy of Sciences, 19B Yuquan Road, Beijing 100049, China}
\address[3]{University of Chinese Academy of Sciences, Beijing 100049, China}
\address[4]{School of Physics and Electronic Science, Guizhou Normal University, Guiyang 550001, China}
\address[5]{Guizhou Provincial Key Laboratory of Radio Astronomy and Data Processing, Guizhou Normal University, Guiyang 550001, China}
\address[6]{School of Physics and Optoelectronics, Xiangtan University, Xiangtan 411105, China}
\address[7]{School of Computer and Information,Dezhou University,Dezhou 253023,Shandong,China}
\address[8]{Institute of Astrophysics, Central China Normal University, Wuhan 430079, China}


\abstract{In the era of time-domain, multi-messenger astronomy, the detection of transient events on the high-energy electromagnetic sky has become more important than ever. 
The Gravitational wave high-energy Electromagnetic Counterpart All-sky Monitor (GECAM) is a dedicated mission to monitor gamma-ray transients, launched in December, 2020. A real-time on-board trigger and location software, using the traditional signal-to-noise ratio (SNR) method for blind search, is constrained to relatively bright signals due to the limitations in on-board computing resources and the need for real-time search. In this work, we developed a ground-based  pipeline for GECAM to search for various transients, especially for weak bursts missed by on-board software. This pipeline includes both automatic and manual mode, offering options for blind search and targeted search.
The targeted search is specifically designed to search for interesting weak bursts, such as gravitational wave-associated gamma-ray bursts (GRBs).
From the ground search of the data in the first year, GECAM has been triggered by 54 GRBs and other transients, including soft gamma-ray repeaters, X-ray binaries, solar flares, terrestrial gamma-ray flashes. 
We report the properties of each type of triggers, such as trigger time and light curves. 
With this search pipeline and assuming a soft Band spectrum, the GRB detection sensitivity of GECAM is increased to about 1.1E-08 erg cm$^{-2}$ s$^{-1}$ (10 keV $-$ 1000 keV, burst duration of 20 s).
These results demonstrate that the GECAM ground search system (both blind search and targeted search) is a versatile pipeline to recover true astrophysical signals which were too weak to be found in the on-board search.}

\keywords{Gamma-ray sources, Gravitational waves, Data analysis, Gamma-ray telescopes}

\PACS{07.85.±m, 95.85.Sz, 07.05.Kf, 95.55.Ka}

\maketitle


\begin{multicols}{2}
\section{Introduction}\label{sec:Introduction}
The three observing runs (O1, O2, O3) \cite{2019PhRvX...9c1040A,2021PhRvX..11b1053A,2023PhRvX..13d1039A} of the Advanced LIGO/Virgo brought groundbreaking discoveries: both the first detection of merging black holes (BHs) \cite{2016PhRvL.116f1102A} and the association of the gravitational wave (GW 170817) \cite{TheLIGOScientific:2017qsa} and the short GRB (GRB 170817A) \cite{Goldstein:2017mmi, Savchenko:2017ffs}, which heralded a new era of multi-messenger astronomy. Since GRB 170817A with low luminosity is considered to be an off-axis short gamma-ray burst (sGRB) \cite{2017ApJ...848L..21A,2018MNRAS.481.1597G}, if more off-axis sGRBs are also associated with GW events, then these sGRBs are expected to be very weak. On the other hand, Observing Run 4 (O4), with a higher sensitivity of 150-180 Mpc \footnote{https://observing.docs.ligo.org/plan/}, started on May 24, 2023 , and will continue until June 9, 2025. During such a sensitive observation, their detected GW events would likely be further and the presumptive gamma-ray bursts associated with those GW events would probably also be very weak.

Recent discovery of a non-thermal X-ray burst from the Galactic magnetar SGR J1935+2154 associated with the fast radio burst (FRB 200428) provides an unambiguous proof that SGR J1935+2154 is the first confirmed source that can radiate the FRB \cite{2020Natur.587...59B,2020Natur.587...54C,2020ApJ...898L..29M,2021NatAs...5..401T,2021NatAs...5..378L,2021NatAs...5..372R}. As most of the detected FRBs are of extragalactic origin (e.g. Lorimer et al. 2007 \cite{2007Sci...318..777L}), if they are also from extragalactic magnetars (e.g. Yang et al. 2020 \cite{2020ApJ...899..106Y}) and associated with X-ray bursts, then these X-ray bursts with short duration could have low luminosity.

The high-sensitivity method to search for GRBs, especially important weak bursts (such as GRBs associated with GWs and X-ray bursts from extragalactic magnetars associated with FRBs), is of great significance in multi-messenger and multi-wavelength astronomy. 
Blackburn et al. developed a coherent method to targeted search the GBM data for transient events in temporal coincidence with GWs \cite{Blackburn:2014rqa}, which has higher sensitivity to search for gamma-ray transients and is widely used in other instruments (e.g. \textit{Swift}-BAT). The idea of this method is to produce an expected count rate from multiple energy channels/detectors and then compare it to the observed counts using a log-likelihood ratio. The log-likelihood ratio quantifies the probability that an astrophysical source is present versus a background-only hypothesis.
In addition to the trigger time, this method can provide a localization \cite{2017ApJ...848L..14G} and help to study the different spectral components \cite{2018ApJ...863L..34B}. Cai et al. improved the coherent search to filter out false triggers caused by spikes in light curves \cite{10.1093/mnras/stab2760} and developed a burst search method based on likelihood ratio in Poisson statistics \cite{2023MNRAS.518.2005C}, which has advantages compared to the Gaussian case, especially for weak bursts.

Gravitational wave burst high-energy Electromagnetic Counterpart All-sky Monitor (GECAM) is a dedicated mission for gamma-ray transients \cite{LiXinQiao2020,2021arXiv211204772L,LiXinQiao2022,2021arXiv211204774A}. There are currently two distinct searches used on GECAM data for gamma-ray transients: One is the real-time on-board trigger and location software for automatic detection of GRBs \cite{2021arXiv211205101Z,HuangYue2022}. It searches for a statistically significant rate increase above the background rate of different energy ranges and timescales based on SNR. The other is ground search (for more details in the following sections) based on the coherent search method, consisting of blind search and targeted search, which is more complicated and sensitive than the in-flight search.

In this paper, we present the methodology and results of the ground burst search of GECAM data for gamma-ray transients. After a brief description of the coherent method for the ground search pipeline, we focus on the results of the ground search during its first year operation (2021, excluding the first month, the early days of the pipeline commissioning operation).
This paper is structured as follows: The GECAM instrument is introduced in section ~\ref{sect:gecam}. The ground search pipeline and results are presented in section \ref{sect:software} and section \ref{sect:Performance}, respectively.  Finally, we give discussion and conclusion in section~\ref{sect:discussion}.

\section{Instrumentation}
\label{sect:gecam}
Gravitational Wave High-energy Electromagnetic Counterpart All-sky Monitor (GECAM)\cite{LiXinQiao2020,2021arXiv211204772L} consists of two microsatellites with a low earth orbit height of 600 kilometers and an inclination angle of 29$^{\circ}$. The payload of each GECAM satellite (i.e., GECAM-A, GECAM-B) is equipped with 25 Gamma-Ray Detectors (GRDs) \cite{2021arXiv211204774A} and 8 Charged Particle Detectors (CPDs) \cite{2021arXiv211205314X}. Thanks to the large FOV (i.e., \textgreater 90 \% of all sky),
highest time resolution (0.1 $\mu$s), broader energy band (10 keV $-$ 5 MeV) and moderate location \cite{2023ApJS..265...17Z} , GECAM is a versatile gamma-ray monitor for gamma-ray bursts (GRBs), soft gamma repeater (SGR) bursts and other transients \cite{stac085,2024SCPMA..6789511Z,2023arXiv230301203A}.

GECAM has three types of science data: the event-by-event data (EVT), binned time data (BTIME) and binned spectrum data (BSPEC). The EVT data consists of individual detector events, each tagged with arrival time (0.1 $\mu$s resolution), energy (4096 raw channels), detector ID, gain type (high or low gain: the detection ranges are about 10-300 keV for high gain and 300 keV-5 MeV for low gain) and dead time \cite{2021arXiv211204786L}. The BTIME data has a finer time resolution of 50 ms and coarse energy resolution (8 channels), while the BSPEC data has higher energy resolution (128 channels) but a more coarse time resolution of 1s.

\section{Ground Search Pipeline}
\label{sect:software}

\begin{figure*}[t]
\centering
\includegraphics[width=1\textwidth]{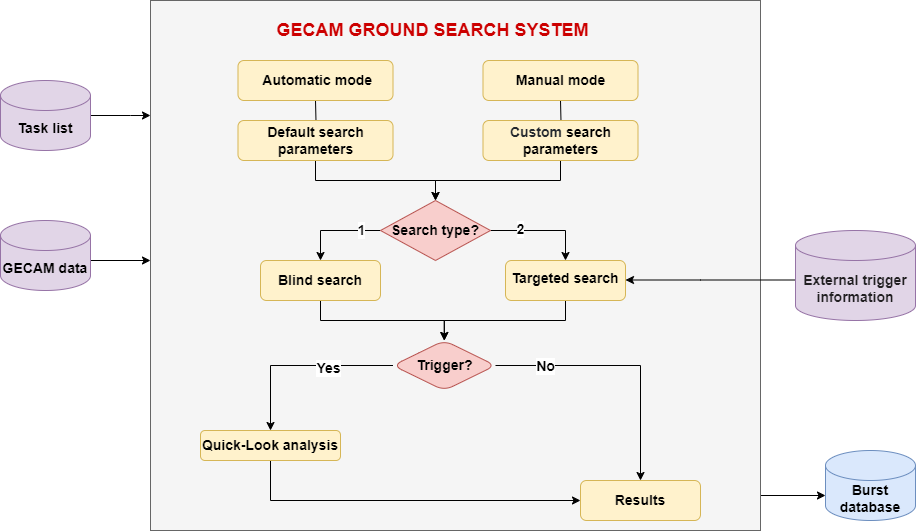}
\caption{Illustration of the framework of the GECAM ground search pipeline.}.
\label{figure_GroundSearchFlow}
\end{figure*}

In order to maximize the GECAM potential for multi-messenger and multi-wavelength observations, the ground search pipeline is developed to identify bursts, especially for important and weak bursts missed by the on-board software. 
The ground pipeline has two search modes: automatic mode and manual mode, both including blind search and targeted search (see Figure \ref{figure_GroundSearchFlow}). To automatically search for transients, the individual processing steps and their dependencies have to be orchestrated in a data analysis flow, that does not need human intervention. This developed pipeline automatically downloads the satellite data as soon as they are available through reading the task list, runs all search analysis steps and uploads a result report to the burst database.

Based on the timing properties of gamma-ray transients and the results of extensive search test using GECAM observed data, we determined the default search parameters (see Table \ref{Tab1}, including search time scale, search phase, search threshold and corresponding false alarm probability) to ensure the effectiveness of the algorithms, which are used for automatic mode. As for manual mode, these parameters could be defined by a burst advocate (a designated individual responsible for analyzing candidate burst events in detail).
The targeted search is a dedicated mode to examine GECAM data for weak counterparts to GW events, FRBs, neutrinos, and other interesting astrophysical signals, so the external trigger information including the trigger time and location of burst is also used as input for the targeted search. 

If multiple triggers are detected simultaneously in the same data file by the ground search, trigger analysis can identify triggers from the same source by comparing their trigger times and define the earliest trigger time among them as the start time of the trigger.

\subsection{Data}
The GECAM real-time in-flight trigger and localization software search BTIME data for triggers, which means that the minimum timescale is 50 ms \cite{2021arXiv211205101Z,HuangYue2022}. The ground-based analysis primarily searches for triggers in both high-precision EVT data and BTIME data, which allows us to exploit some extreme short bursts with durations shorter than 50 ms. 

\begin{table*}[t]
\begin{center}
\caption[]{Trigger algorithms used in the ground search and algorithms statistics $^{1}$}\label{Tab1}
\begin{threeparttable}
\begin{tabular}{ccccccccccc}
\hline\noalign{\smallskip}
\hline\noalign{\smallskip}
IDs & Timescale (s) & Phase (s) & Threshold (LR) & FAP$^{2}$  & GRBs & SGRs & SFs & TGFs & CPs & Others$^{3}$\\
  \hline\noalign{\smallskip}
1 & 0.05 & 0 & 23.62 & $6.28 \times 10^{-12}$ & 5 & 42 & 0 & 8 &18 & 40\\

\hline
2 & 0.1 & 0 & 22.94 & $1.26 \times 10^{-11}$  & 9 & 17 & 0 & 0 & 13 & 23\\
3 & 0.1 & 0.05 & 22.94 & $1.26 \times 10^{-11}$ &   \\
\hline

4 & 0.2 & 0 & 22.26 & $2.51 \times 10^{-11}$ & 8 & 4 & 0 & 0 & 10 & 4\\
5 & 0.2 & 0.1 & 22.26 & $2.51 \times 10^{-11}$ &  \\
\hline

6 & 0.5 & 0 & 21.36 & $6.28 \times 10^{-11}$ & 11 & 0 & 4 & 0 &13 & 11\\
7 & 0.5 & 0.25 & 21.36 & $6.28 \times 10^{-11}$ &  \\
\hline

8 & 1 & 0 & 20.68 & $1.26 \times 10^{-10}$ &  5 & 0 & 16 & 0 & 23 & 27\\
9 & 1 & 0.5 & 20.68 & $1.26 \times 10^{-10}$ &  \\
\hline

10 & 2 & 0 & 20.01 & $2.51 \times 10^{-10}$ & 11 & 0 & 36 & 0 &74 & 87\\
11 & 2 & 1 & 20.01 & $2.51 \times 10^{-10}$ &  \\
\hline

12 & 4 & 0 & 19.33 & $5.02 \times 10^{-10}$ & 5 & 0 & 47 & 0 & 227 & 141\\
13 & 4 & 2 & 19.33 & $5.02 \times 10^{-10}$ &  \\

\noalign{\smallskip}\hline
\end{tabular}

\begin{tablenotes}
    \footnotesize
    \item[1] The number of events triggered by each criterion (i.e., each search timescale) is provided. For example, 9 GRBs were triggered by algorithms 2 or 3 with a timescale of 0.1 s.
    \item[2] False alarm probability is calculated based on Wilks’ theorem.
    \item[3] Other sources include X-ray binaries, Earth occultation of known sources (e.g. Sco X-1), instrument effects and uncertain triggers.
\end{tablenotes}
\end{threeparttable}    
\end{center}
\end{table*}

\subsection{Methodology}

The ground sensitive coherent search method is widely performed and improved to search for GRBs and magnetar bursts \cite{Blackburn:2014rqa,2021GCN.30125....1F,2021GCN.30140....1C,Goldstein:2016zfh,Goldstein:2019tfz,10.1093/mnras/stab2760}. 
Data from the 25 detectors of GECAM-B are processed coherently to achieve a greater sensitivity to weak signals. Three Band spectra \cite{Band:1993} representing spectrally hard, normal, and soft GRBs (Table \ref{tab:bandfunction}) are folded through the direction-depended detector responses to produce templates of expected counts.
The templates are then compared to the observed counts in each channel of each detector via a log-likelihood ratio.

\begin{table}[H]
\begin{center}
\caption{Model Band Spectral Parameters}\label{tab:bandfunction}
\begin{tabular}{  p{1.5cm}  p{1.5cm}  p{1.5cm} p{1.5cm}} 
\hline
    Spectrum &  $\alpha$  &  $\beta$ &   \Ep\\
    \hline
    soft   &  -1.9   & -3.7  &  70 keV \\
 
    normal & -1      & -2.3  &  230 keV\\

    hard   & 0       & -1.5  &  1 MeV\\
    \hline 
\end{tabular}
\end{center}
\end{table}

Following the presentation in Blackburn et al. \cite{Blackburn:2014rqa} and Cai et al. \cite{10.1093/mnras/stab2760}, the log-likelihood ratio of coherent search method could be formulated as:

\begin{equation}
{{P}({d}_{i}|{H}_{1})}= \prod_{i=1}^j \frac{1}{\sqrt{2\pi}{\sigma}_{{d}_{i}}}\rm exp(-\frac{({\widetilde{d}_{i}-{r}_{i}s)^2}}{2\sigma^{2}_{{d}_{i}}}) \,\,(i=1,2...j), \label{eq1}
\end{equation}

\begin{equation}
{{P}({d}_{i}|{H}_{0})}= \prod_{i=1}^j \frac{1}{\sqrt{2\pi}{\sigma}_{{n}_{i}}}\rm exp(-\frac{{\widetilde{d}_{i}}^2}{2\sigma^{2}_{{n}_{i}}}), \label{eq2}
\end{equation}

where $i$ is the number of data sets in each channel and detector, $j$ is the total number of detectors and channels, 
${d}_{i}$ is the observed data (counts) and ${\sigma}_{{d}_{i}}$ is the standard deviation of the expected data (background+signal), ${n}_{i}$ is the estimated background and ${\sigma}_{{n}_{i}}$ is the standard deviation of the background data, ${r}_{i}$ and $s$ represent the instrument response and the intrinsic source amplitude, respectively.

\begin{equation}
\widetilde{d}_{i}={d}_{i}-\left\langle{n}_{i}\right\rangle,
\end{equation}

where $\widetilde{d}_{i}$ is the background-subtracted data (i.e. net counts).

Then the log-likelihood ratio (hereafter LR) is defined as:
\begin{equation}
{ \mathcal L} = {\rm ln} \frac{{P}({d}_{i}|{H}_{1})}{{P}({d}_{i}|{H}_{0})} = \sum_{i=1}^j[{\rm ln}\frac{{\sigma}_{{n}_{i}}}{{\sigma}_{{d}_{i}}} + \frac{{\widetilde{d}_{i}}^2}{2\sigma^{2}_{{n}_{i}}} - \frac{({\widetilde{d}_{i}-{r}_{i}s)^2}}{2\sigma^{2}_{{d}_{i}}}].   \label{eq3}
\end{equation}

where $\mathcal L$ represents LR. 

We estimate the background of each detector by fitting a polynomial to data from $T_{\rm \,0}$-15 s to $T_{\rm \,0}$+15 s, excluding the interval form $T_{\rm \,0}$-5 s to $T_{\rm \,0}$+5 s. And a separate fit is done for each detector/energy channel combination.

The dependence of the response factors on sky location is
complicated, so the likelihood ratio is calculated over a sample
grid of all possible locations. For blind search,  since the location is unknown, we divided the whole sky into 3072 pixels with the same surface area using Hierarchical Equal Area and iso-Latitude Pixelation (HEALPIX). As for targeted search, we can calculate the likelihood ratio using the location of targeted source observed by other instruments (e.g., \textit{Insight}-HXMT \cite{2020SCPMA..6349502Z}, \textit{Swift}/BAT \cite{2013ApJS..209...14K}, \textit{Fermi}/GBM \cite{2009ApJ...702..791M}, LIGO \cite{1992Sci...256..325A}, Virgo \cite{2012JInst...7.3012A}, and KAGRA \cite{2021PTEP.2021eA103A}).

As mentioned above, for the blind search, the expected counts are estimated at 3072 positions across the entire sky, leading to a significant increase in computational complexity. To overcome this, the code is developed with Python 3 and optimized using the multiprocessing package. The position corresponding to the maximum LR is considered to be the most possible location of the trigger. 
The preliminary location of the trigger provided by ground search, can help classify this candidate. In addition, a dedicated location pipeline has been performed with a smaller localization error \cite{2023ApJS..265...17Z}.

\subsection{Quick-Look Analysis}
Quick-look data including trigger time, significance and location, is generated when a candidate triggers the ground search. Also the light curves of different detectors and energy ranges, sky map, geographic location map of the satellite and scatter plot of events are generated.

The GECAM team assigns a burst advocate (BA) to inspect the quick-look data promptly and perform additional analysis as appropriate to identify the real astrophysical signal. Normally, the in-flight and on-ground auto-analysis is reviewed by BA, in consultation with other team members, and may be corrected based on the additional analysis and information from other instruments (e.g., \textit{Insight}-HXMT, \textit{Fermi}/GBM, \textit{Swift}/BAT, LIGO, Virgo, and KAGRA).

\section{Search Results}
\label{sect:Performance}

\subsection{Blind Search}
\subsubsection{Trigger statistics}

During the first year (2021) of the operation, excluding the first month of 2021 (the early days of the pipeline commissioning operation), 54 triggers are classified as GRBs by BA, 12 of which did not trigger the on-board search \cite{HuangYue2022}. In addition, the number of SGRs and solar flares detected by the ground search is higher than the on-board search (see Table \ref{Tab2}). The observation time lengths per month are also listed in Table \ref{Tab2}.

\begin{table*}[t]
\scriptsize
\centering
\caption{The results of on-board search and ground search for GRBs, SGRs and SFs.} \label{Tab2}
\begin{threeparttable}
\begin{tabular}{ccccccccc}
\toprule
\multirow{2}{*}{Year-Month} & \multirow{2}{*}{Observation time length (ks)$^{1}$} & \multirow{2}{*}{proportion$^{2}$} & \multicolumn{2}{c}{Gamma Ray Bursts} & 
\multicolumn{2}{c}{Soft Gamma Repeaters} & \multicolumn{2}{c}{Solar Flares} \\
\cmidrule(r){4-5} \cmidrule(r){6-7} \cmidrule(r){8-9}
& & & \multicolumn{1}{c}{on-board} & \multicolumn{1}{c}{ground}
& \multicolumn{1}{c}{on-board} & \multicolumn{1}{c}{ground}
& \multicolumn{1}{c}{on-board} & \multicolumn{1}{c}{ground} \\
\midrule
2021-02	&	1209.127 	&	0.500 	&	4	&	4	&	1	&	2	&	2	&	2	\\
2021-03	&	1182.395 	&	0.441 	&	4	&	5	&	0	&	0	&	2	&	4	\\
2021-04	&	890.333 	&	0.343 	&	4	&	7	&	0	&	0	&	9	&	9	\\
2021-05	&	1068.472 	&	0.399 	&	4	&	5	&	0	&	0	&	15	&	23	\\
2021-06	&	992.670 	&	0.383 	&	5	&	5	&	0	&	0	&	2	&	2	\\
2021-07	&	778.319 	&	0.291 	&	1	&	1	&	6	&	6	&	4	&	12	\\
2021-08	&	959.727 	&	0.358 	&	2	&	2	&	0	&	0	&	5	&	5	\\
2021-09	&	868.522 	&	0.335 	&	3	&	5	&	22	&	52	&	0	&	0	\\
2021-10	&	1103.470 	&	0.412 	&	1	&	2	&	2	&	2	&	19	&	31	\\
2021-11	&	1035.975 	&	0.400 	&	5	&	7	&	0	&	1	&	2	&	2	\\
2021-12	&	1431.131 	&	0.534 	&	9	&	11	&	0	&	0	&	13	&	13	\\
\bottomrule
\end{tabular}
\begin{tablenotes}
    \footnotesize
    \item[1] Observation time length excluding the first month with the early days of the pipeline commissioning operation.
    \item[2] The proportion of the observation time length of GECAM-B in a month.
    \end{tablenotes}
    \end{threeparttable}
\end{table*}

\begin{figure}[H]
		\centering
		\includegraphics[width=0.95\columnwidth]{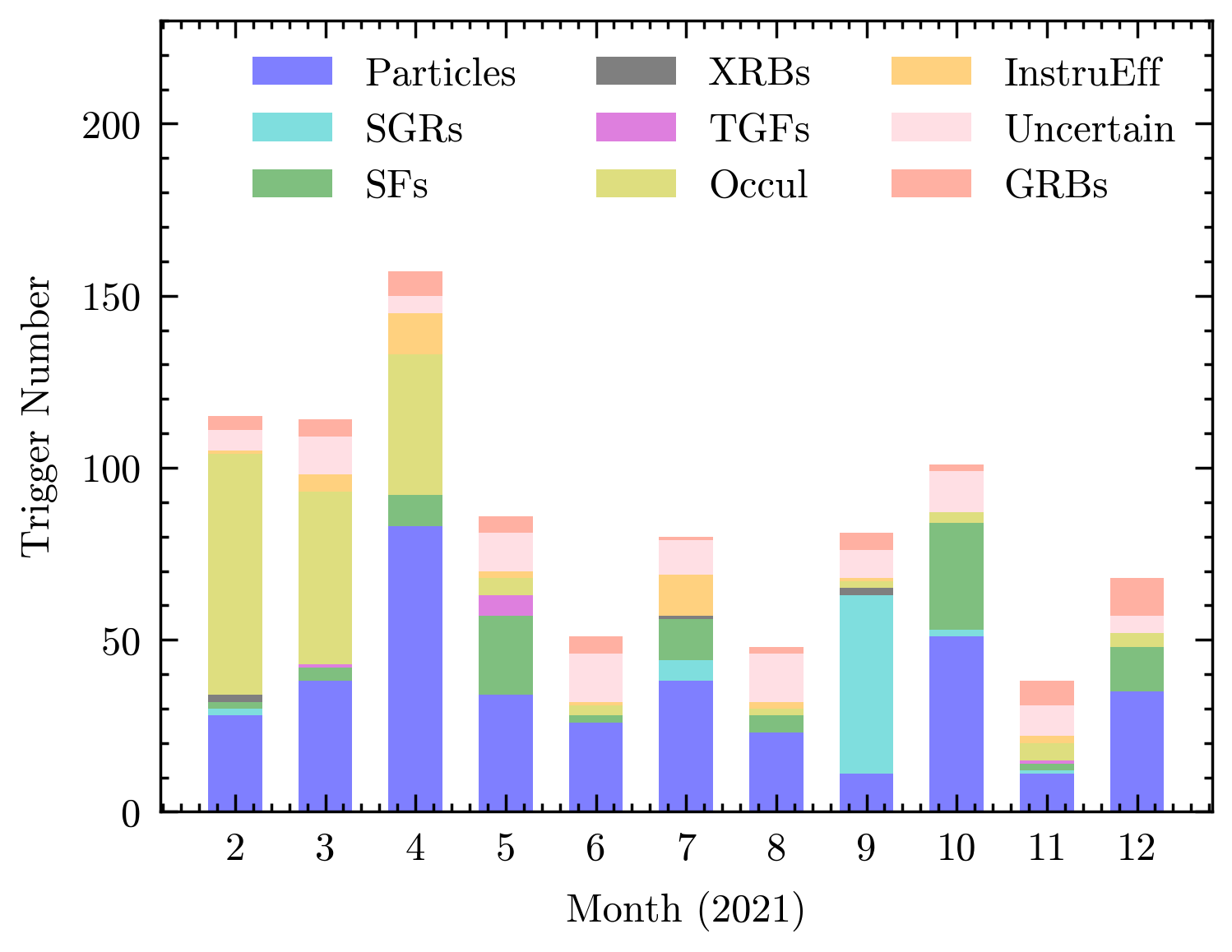}
		\caption{Monthly ground trigger statistics for 2021 (from February to December). Different colors represent different sources: gamma-ray bursts, soft gamma-ray repeaters, solar flares, X-ray binaries, terrestrial-gamma flashes, Earth occultation, particles events, instrument effects and uncertain triggers. Uncertain triggers are too weak to identify without observation of other instruments. We leave the detailed classification of these uncertain triggers only to the future work.}
		\label{figure_TriggerStatistics}
\end{figure}

\begin{figure}[H]
		\centering
		\begin{tabular}{c}
		\includegraphics[width=0.95\columnwidth]{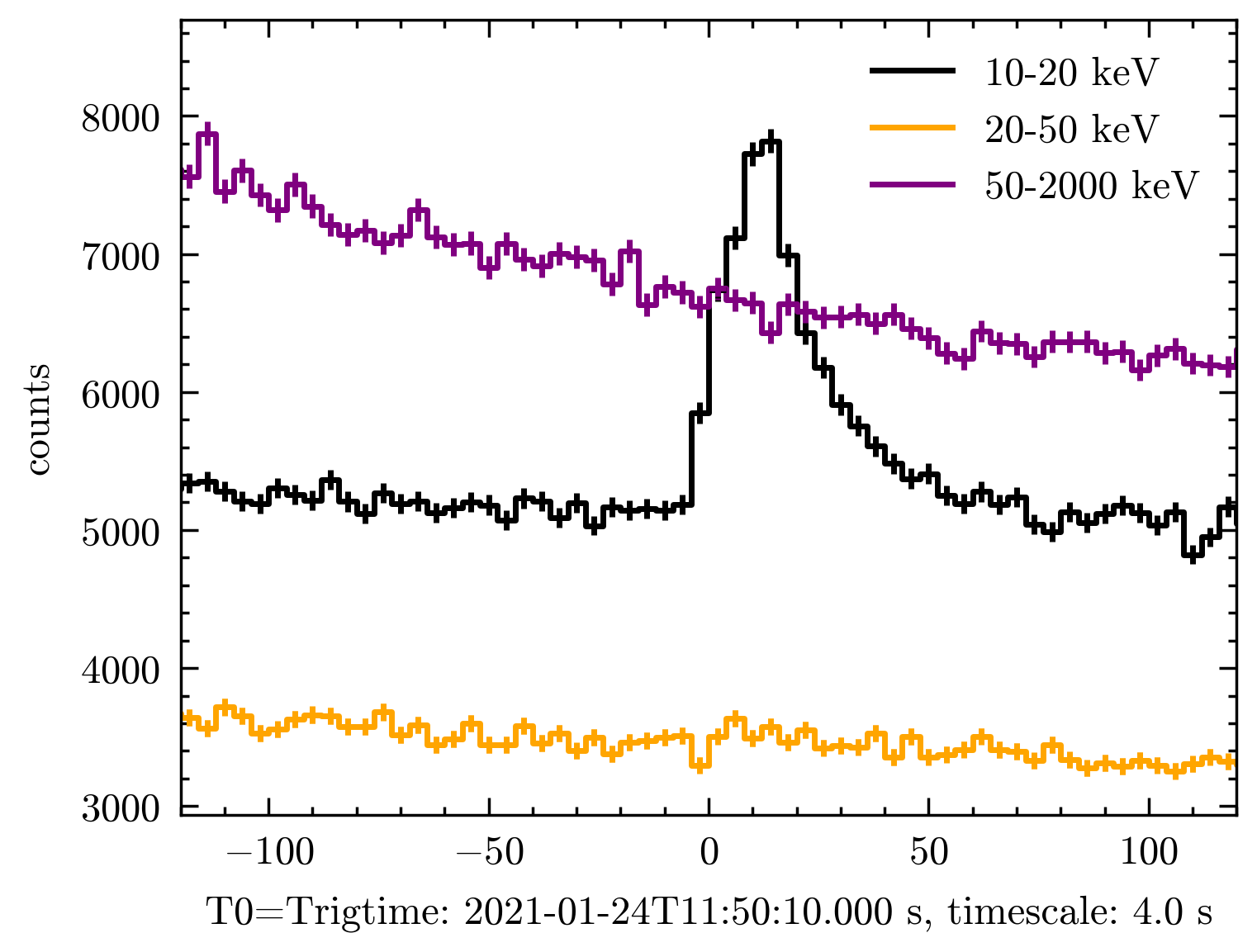} 
		\end{tabular}
		\caption{Light curves of an X-ray binary (4U 0614+09) with different energy ranges (i.e. 10--20 keV, 20--50 keV and 50--2000 keV). Data from GECAM GRD detector \#11, \#12, \#21, \#22.}
		\label{figure_4U 0614+09}
\end{figure}

All triggers are classified as GRBs, SGRs, Solar flares, X-ray binaries, terrestrial gamma-ray flashes, Earth occultation of known sources (e.g. Sco X-1), particles, instrument effects, false triggers, and uncertain triggers (not easy to identify). Table \ref{Tab1} summarizes which trigger algorithm has triggered first on bursts from the different object classes. GRBs triggered all types of algorithms, while SGRs and TGFs tend to trigger shorter timescale algorithms, SFs and most CPs are more likely to trigger longer timescale algorithms.

The statistics of monthly ground trigger (excluding false triggers) is shown in Figure \ref{figure_TriggerStatistics}. The triggers from the instrument effect are mainly caused by the optimization and update of the temperature-bias voltage adjustment parameters \cite{2021arXiv211204772L}. Some interesting candidates (e.g., the candidates of GRBs, SGRs, X-ray binaries and terrestrial gamma-ray flashes) are unable to be confirmed with quick-look analysis due to the weak signals. Detailed analysis of these bursts will be published in the forthcoming burst catalogue paper (Zhang Y. Q. et al., in prep).

\begin{figure}[H]
		\centering
		\begin{tabular}{c}
		\includegraphics[width=0.95\columnwidth]{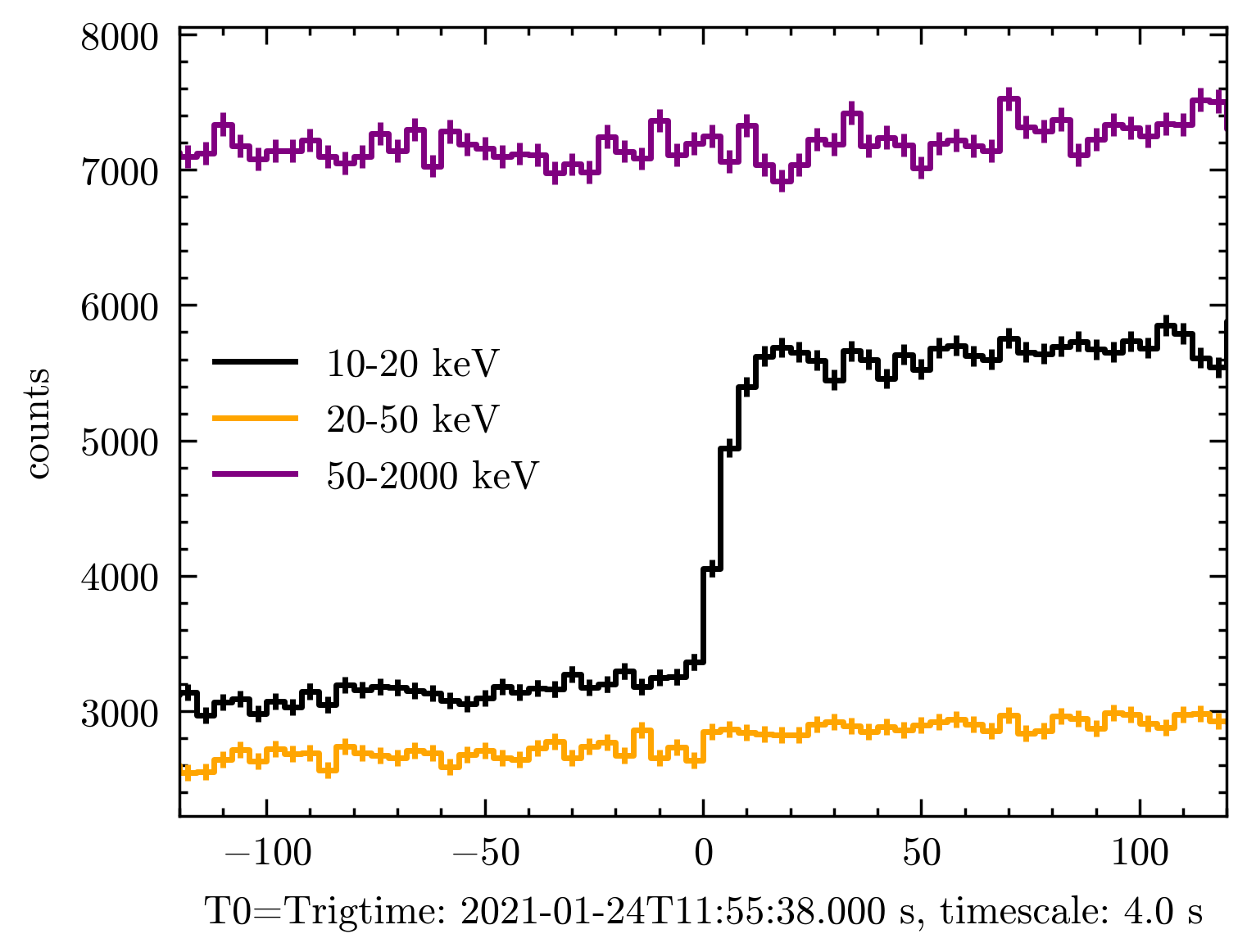} \\
		\end{tabular}
		\caption{Light curves for Earth occultation of Sco X–1 with different energy ranges (i.e. 10--20 keV, 20--50 keV and 50--2000 keV). Data from GECAM GRD detector \#15, \#16, \#17 and \#25.}
		\label{figure_Occultation}
\end{figure}

\subsubsection{Diverse Triggers}
X-ray binary (XRB) is a system with a central compact object (black hole or neutron star) accreting matter from its companion star \cite{1989A&A...225...79H}. Based on the light curve and location, a few triggers are classified as XRB bursts \cite{2021arXiv211204790C}. The light curves of 4U 0614+09 observed by GECAM are shown in Figure \ref{figure_4U 0614+09}, which exhibit a duration of about 50 seconds and show that the increase of the count rate is mainly in the energy range of 10--20 keV.

Step-like occultation features will be observed in its counting rate when a point source of gamma rays (e.g., Sco X-1) crosses the limb of the Earth \cite{2023ApJS..264....5X}. As shown in Figure \ref{figure_Occultation}, the increase in the count rate of Sco X-1 is apparent in the energy range of 10$-$20 keV, because of its soft spectra. 
A significant reduction in the number of Earth occultation of sources over time(e.g., Sco X–1) (see Figure \ref{figure_TriggerStatistics}) is due to the raising of the low-energy threshold of GRD. 

\begin{figure}[H]
		\centering
		\begin{tabular}{c}
	 \includegraphics[width=0.95\columnwidth]{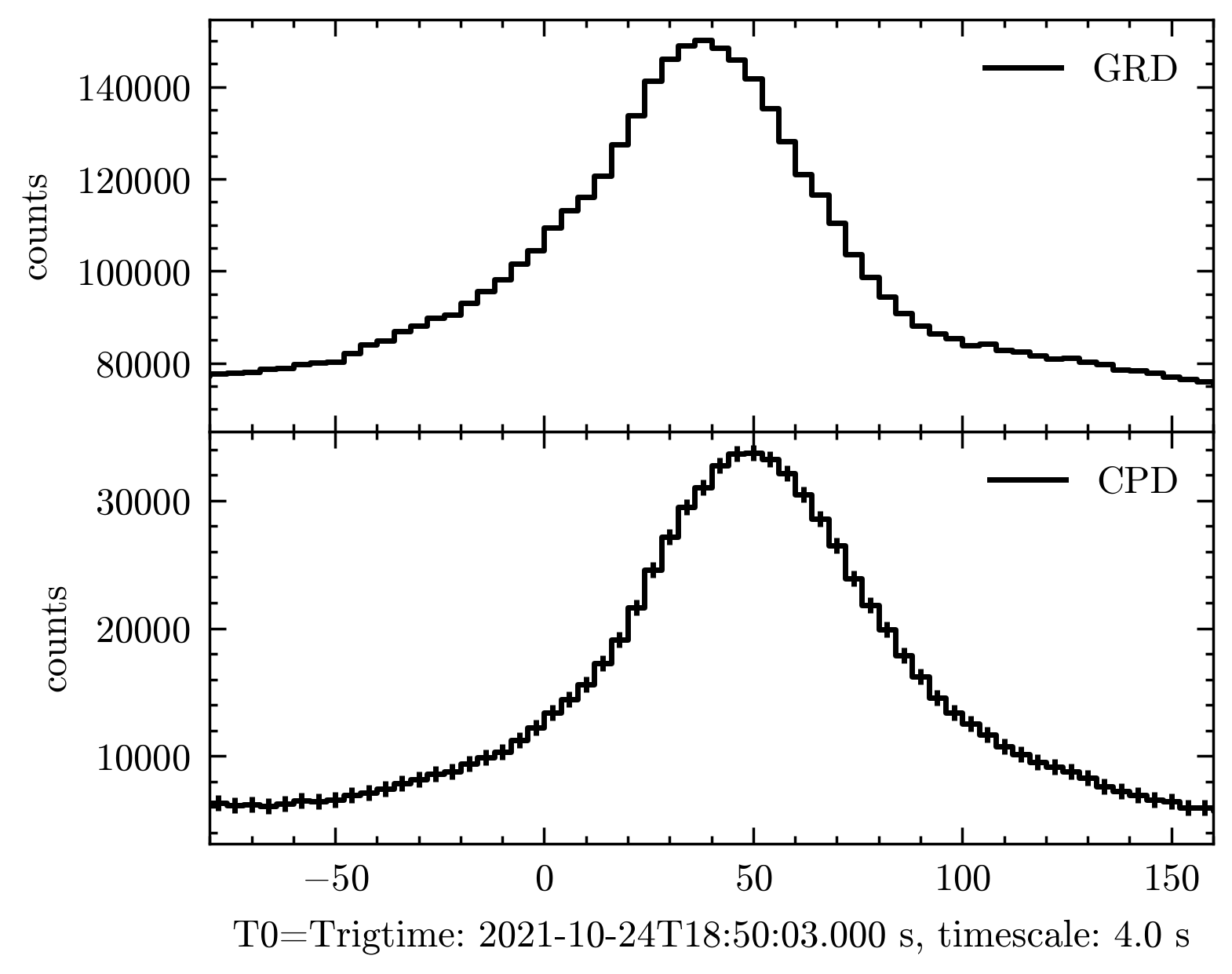} 
		\end{tabular}
		\caption{The light curves of a particle event. \textit{Top}: Data from 25 gamma ray detectors (20$-$2000 keV). \textit{Bottom}: Data from 8 charged particle detectors (\textgreater 100 keV).}
		\label{figure_ParticlesLightCurve}
\end{figure}

We take the light curves of GRDs and CPDs as examples to show a particle event with the trigger time of 2021-10-24T18:50:03.000 UTC (see Figure \ref{figure_ParticlesLightCurve}). The count ratio between CPDs and GRDs is used to identify particle events. Furthermore, particle events occur predominantly in trapped particle regions. Therefore, we also use the geographic location (longitude and latitude) of satellite to identify and remove particles from the burst candidates \cite{HuangYue2022}.

\begin{figure}[H]
		\centering
		\begin{tabular}{c}
		\includegraphics[width=0.95\columnwidth]{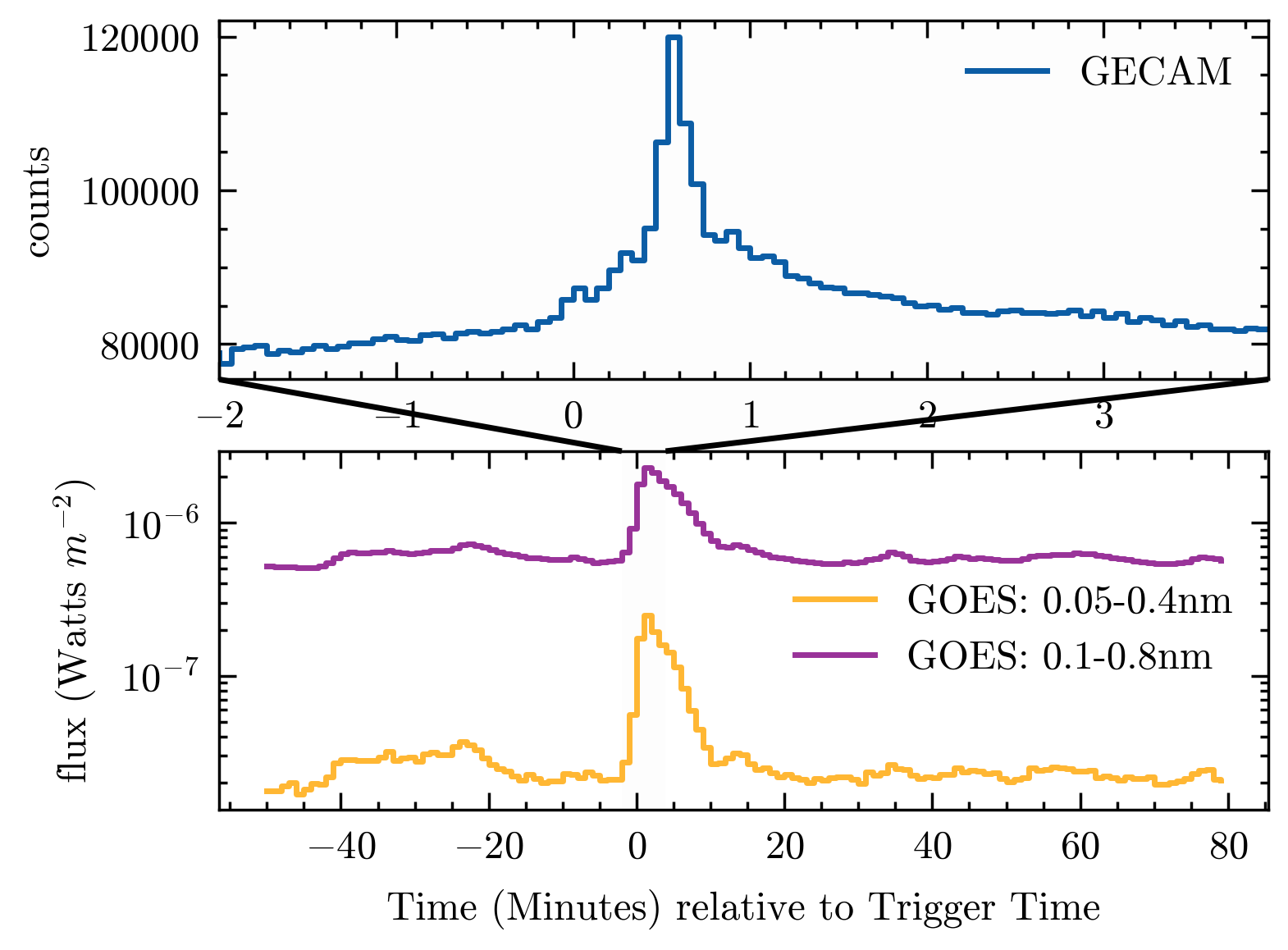} 
		\end{tabular}
		\caption{The light curves of a solar flare. In the top panel, the blue line is the light curve of GECAM GRD data with the bin width of 4 s and energy range of 20$-$2000 keV. The bottom panel is the light curves of GOES data with the bin width of one minute. $T_0$ is the ground trigger time of 2022-05-07T15:05:39.000 UTC. The purple line and orange line are the light curves with 0.1$-$0.8 nm and 0.05$-$0.4 nm, respectively.}
		\label{figure_SolarLightCurve}
\end{figure}

The Solar flare is an intense burst of radiation, which comes from the release of magnetic energy associated with sunspots \cite{2023SCPMA..6659611Z}. As shown in Figure \ref{figure_SolarLightCurve}, a solar flare is detected by GECAM, of which the trigger time is consistent with that of the available Geostationary Operational Environmental Satellite (GOES) data \footnote{https://services.swpc.noaa.gov/json/goes/primary/}. Also we find that the location of this trigger is in agreement with that of the Sun.

GECAM ground search has higher sensitivity than on-board search, which can provide an expanded GRB sample. Among the 54 GRBs detected by ground search, 42 GRBs were also found by on-board search. The remaining 12 GRBs did not trigger the detection threshold in the on-board search (see Table \ref{Tab2} and Table\ref{Tab4}).

\begin{figure*}[t]
		\centering
		\begin{tabular}{c}
		\includegraphics[width=2.0\columnwidth]{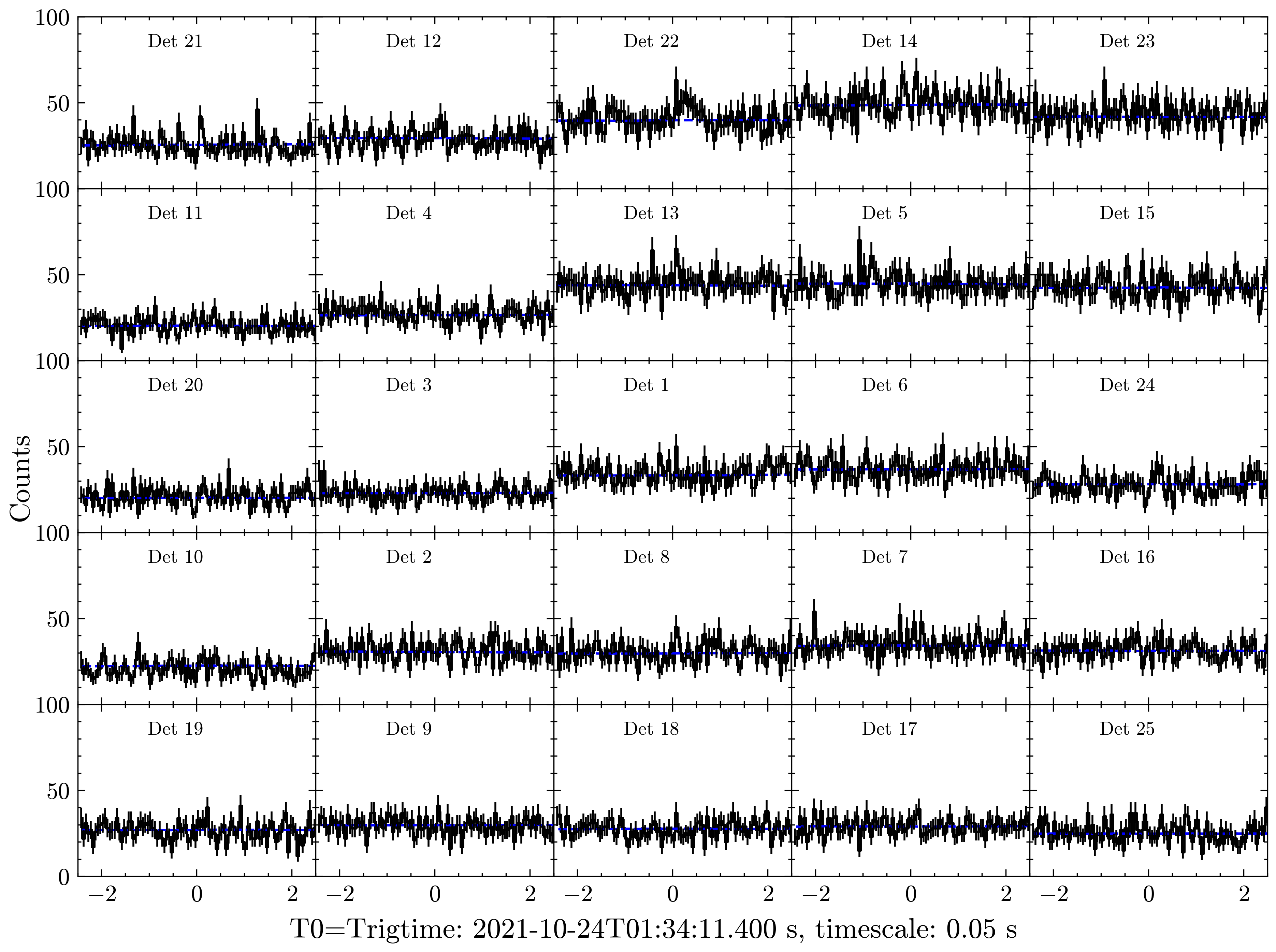} 
		\end{tabular}
		\caption{GECAM GRD light curves of GRB 211024A with the energy range of 20$-$2000 keV, which are distributed according to the position of the detectors installed on the satellite. The horizontal blue dash lines are the estimated background.}
		\label{figure_211024A_1}
\end{figure*}

\begin{figure}[H]
		\centering
		\begin{tabular}{c}
		\includegraphics[width=0.95\columnwidth]{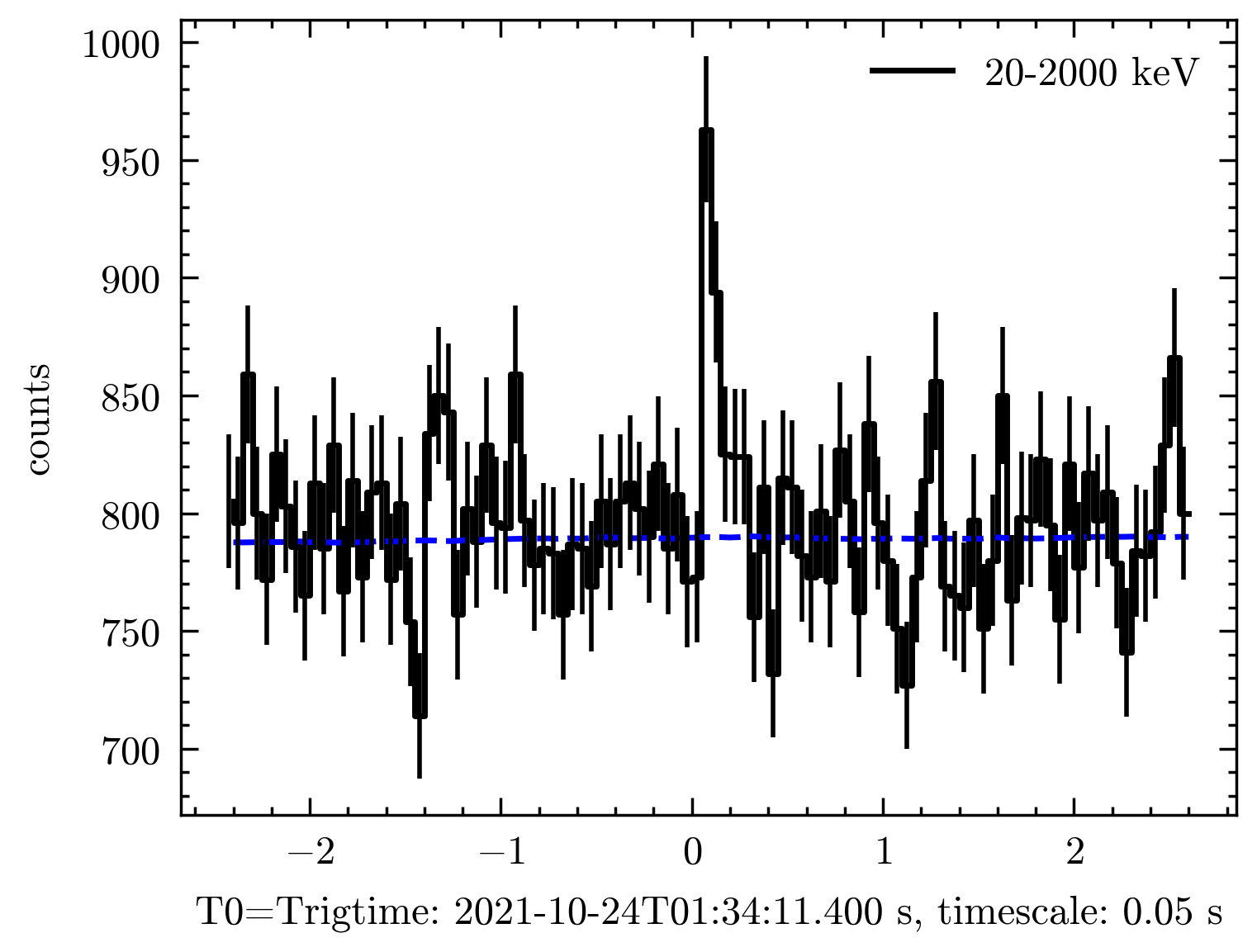} 
		\end{tabular}
		\caption{GECAM all 25 GRD light curves of GRB 211024A  with the trigger time of 2021-10-24T01:34:11.400 UTC and the energy range of 20$-$2000 keV. The horizontal blue dash lines are the estimated background.}
		\label{figure_211024A_2}
\end{figure}

\begin{figure*}[t]
		\centering
		\begin{tabular}{c}
		\includegraphics[width=2.0\columnwidth]{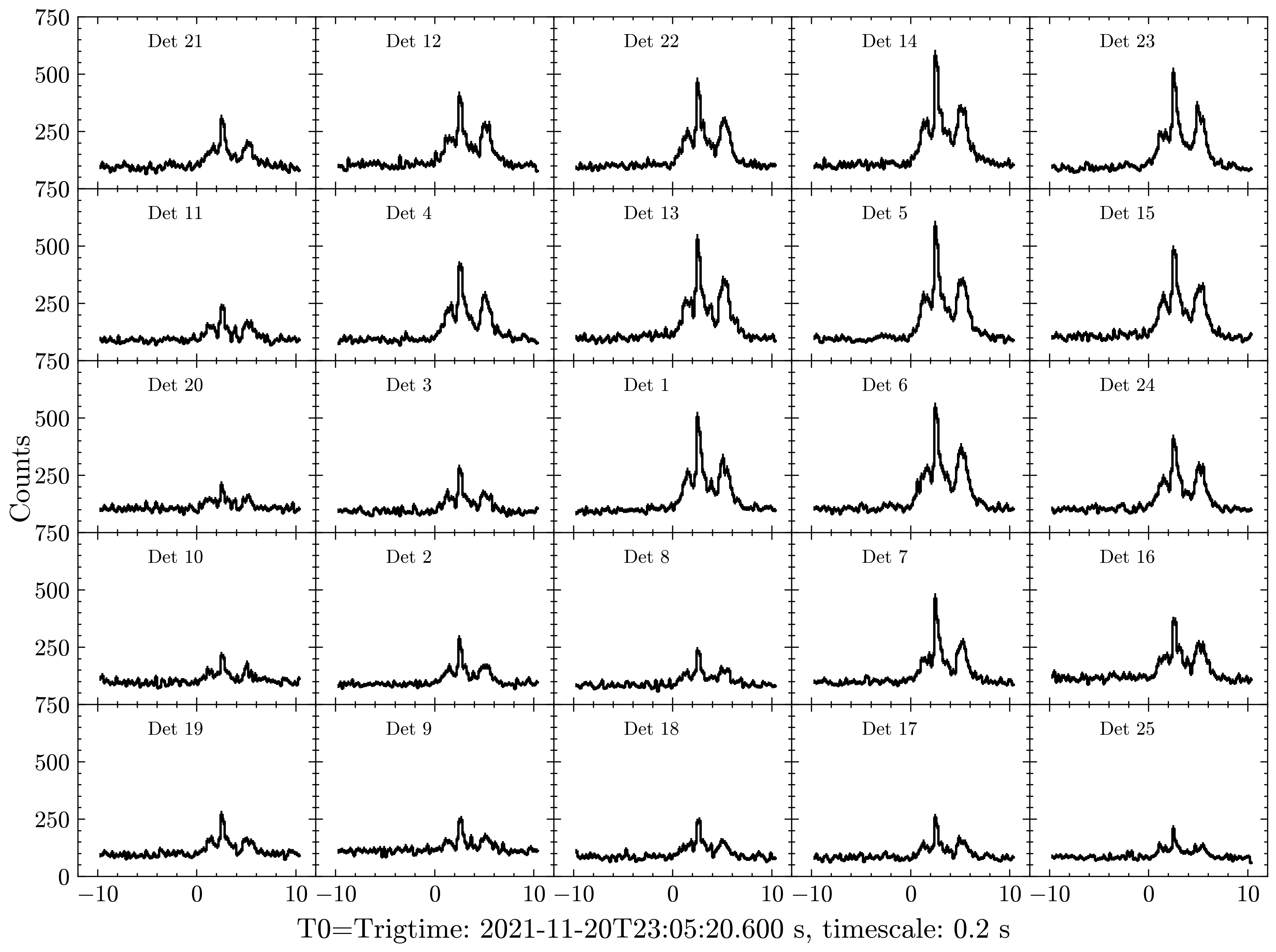} 
		\end{tabular}
		\caption{GECAM GRD light curves of GRB 211120A with the energy range of 20$-$2000 keV, which are distributed according to the position of the detectors installed on the satellite.}
		\label{figure_GRB_211120A_1}
\end{figure*}

\begin{figure}[H]
		\centering
		\begin{tabular}{c}
		\includegraphics[width=0.95\columnwidth]{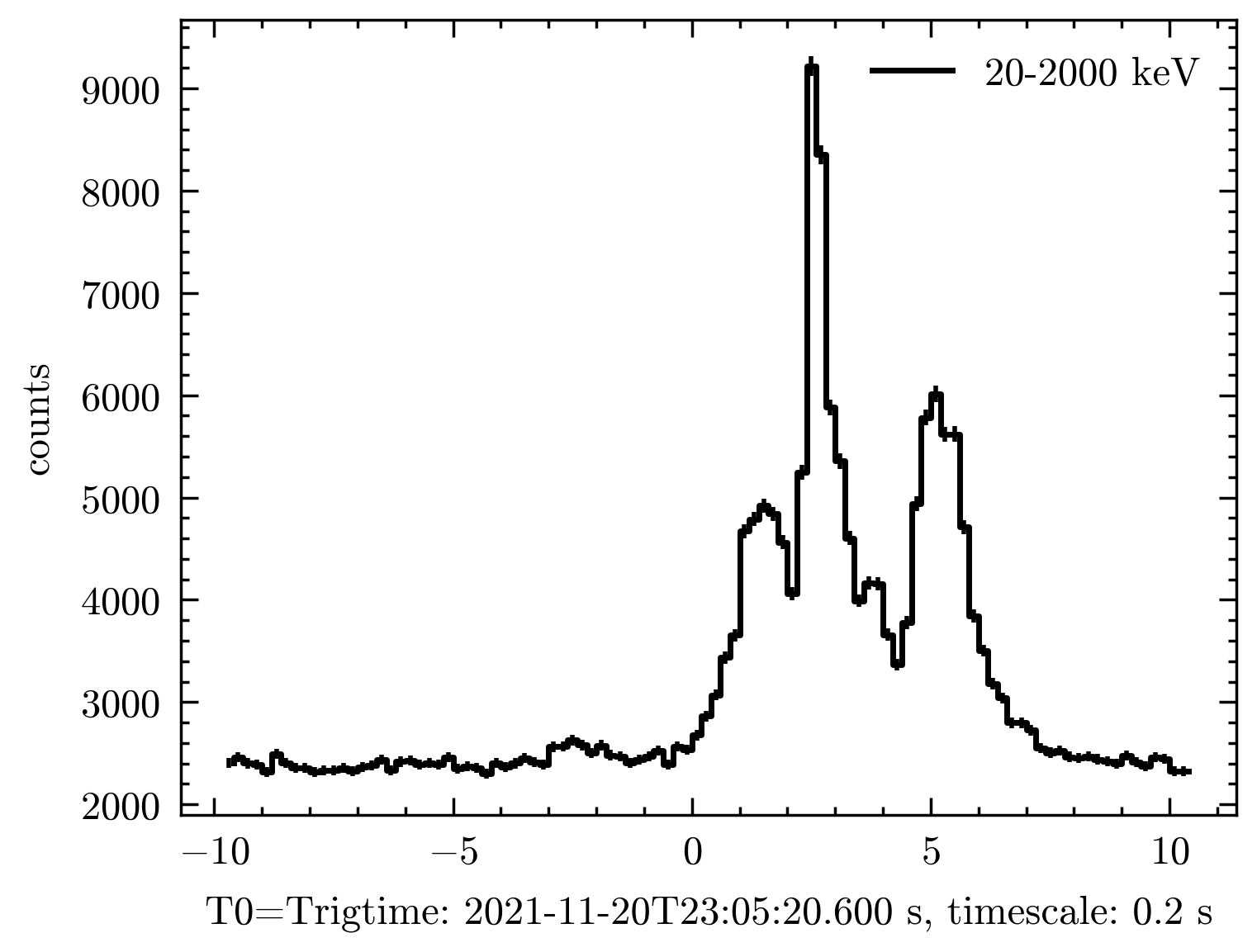} 
		\end{tabular}
		\caption{GECAM all 25 GRD light curves of GRB 211120A with the trigger time of 2021-11-20T23:05:20.600 UTC and the energy range of 20$-$2000 keV.}
		\label{figure_GRB_211120A_2}
\end{figure}

GRB 211024A is detected by both GECAM ground search and \textit{Fermi}/GBM \cite{2021GCN.31011....1P}, but not  triggered the GECAM on-board search. The light curves of each GECAM detector for this burst is shown in Figure \ref{figure_211024A_1}. There are less signals with each detector, while we found a relatively significant signal when all counts of each detector unit are summed together for this burst (see Figure \ref{figure_211024A_2}). According to the on-orbit default parameters, the on-board search trigger is achieved when more than 3 detectors having significance higher than 4 sigma. Therefore, these weak bursts (e.g. GRB 211024A) with less than 3 detectors having significance higher than 4 sigma, can not be found in on-board search. GRB 211120A is a bright long burst (see Figure \ref{figure_GRB_211120A_1} and Figure \ref{figure_GRB_211120A_2}), which were found by on-board search and ground search.

After checking the light curves of both the ground triggered and on-board untriggered 12 GRB bursts, we find that these weak bursts with less background-subtracted counts only can be recovered by the ground search. The properties of these GRBs, including their trigger time, duration, spectral parameters, peak fluxes of different timescales, and fluence, will be reported in the forthcoming burst catalogue paper (Zhang Y. Q. et al., in prep).

\begin{figure}[H]
		\centering
		\begin{tabular}{c}
		\includegraphics[width=0.95\columnwidth]{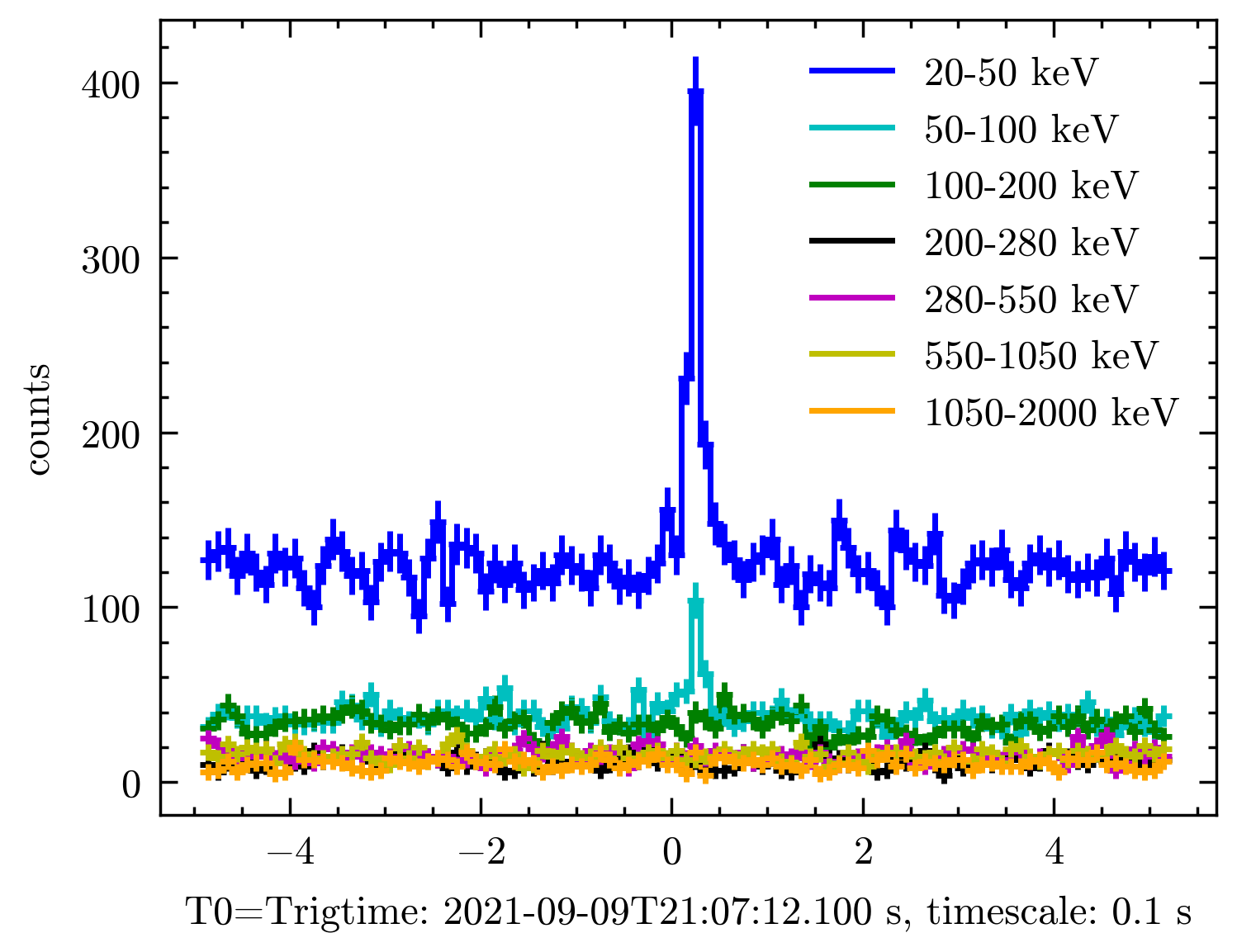} 
		\end{tabular}
		\caption{Light curves of a SGR J1935+2154 burst with different energy ranges. Data from GECAM GRD detector \#5, \#12, \#13, \#22, \#23 and \#24.}
		\label{figure_SGRLightCurve}
\end{figure}

\begin{figure}[H]
		\centering
		\begin{tabular}{c}
		\includegraphics[width=0.95\columnwidth]{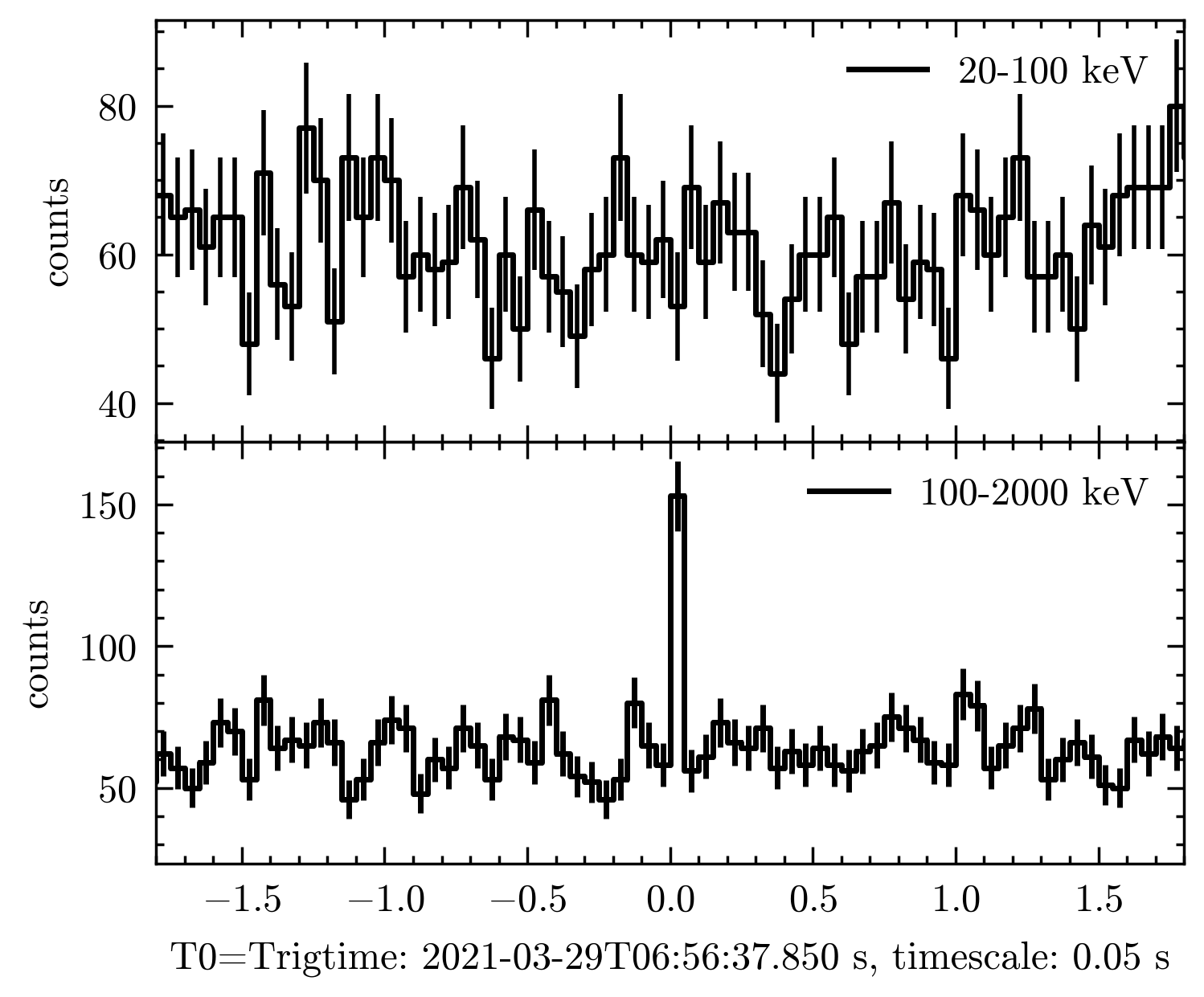} 
		\end{tabular}
		\caption{Light curves of a TGF with the energy range of 20--100 keV and 100--2000 keV, respectively.
  Data from GECAM GRD detector \#5, \#12, \#20, \#21 and \#23.}
		\label{figure_TGFLightCurve1}
\end{figure}

The typical duration of short bursts from SGRs ranges from 0.1 s to 1 s \cite{2015ApJS..218...11C,Lin_2020b}. Figure \ref{figure_SGRLightCurve} shows the light curves of the SGR J1935+2154 burst with a timescale of 0.1 s. A series of bursts from this source triggered GECAM ground search in July and September of 2021 (see Figure \ref{figure_TriggerStatistics}). Recently, a new magnetar, SGR J1555.2-5402 \cite{2021GCN.30120....1P} emitted many short X-ray bursts and triggered a series of astronomical satellites. Most of these bursts triggered GECAM on-board search and ground blind search, while weaker and shorter bursts were found by targeted search (see section \ref{target_search} for more details). The properties of these SGR bursts, including their trigger time, duration and spectra are reported in the burst catalogue paper \cite{2023arXiv230701010X}.

\begin{figure}[H]
		\centering
		\begin{tabular}{c}
		\includegraphics[width=0.95\columnwidth]{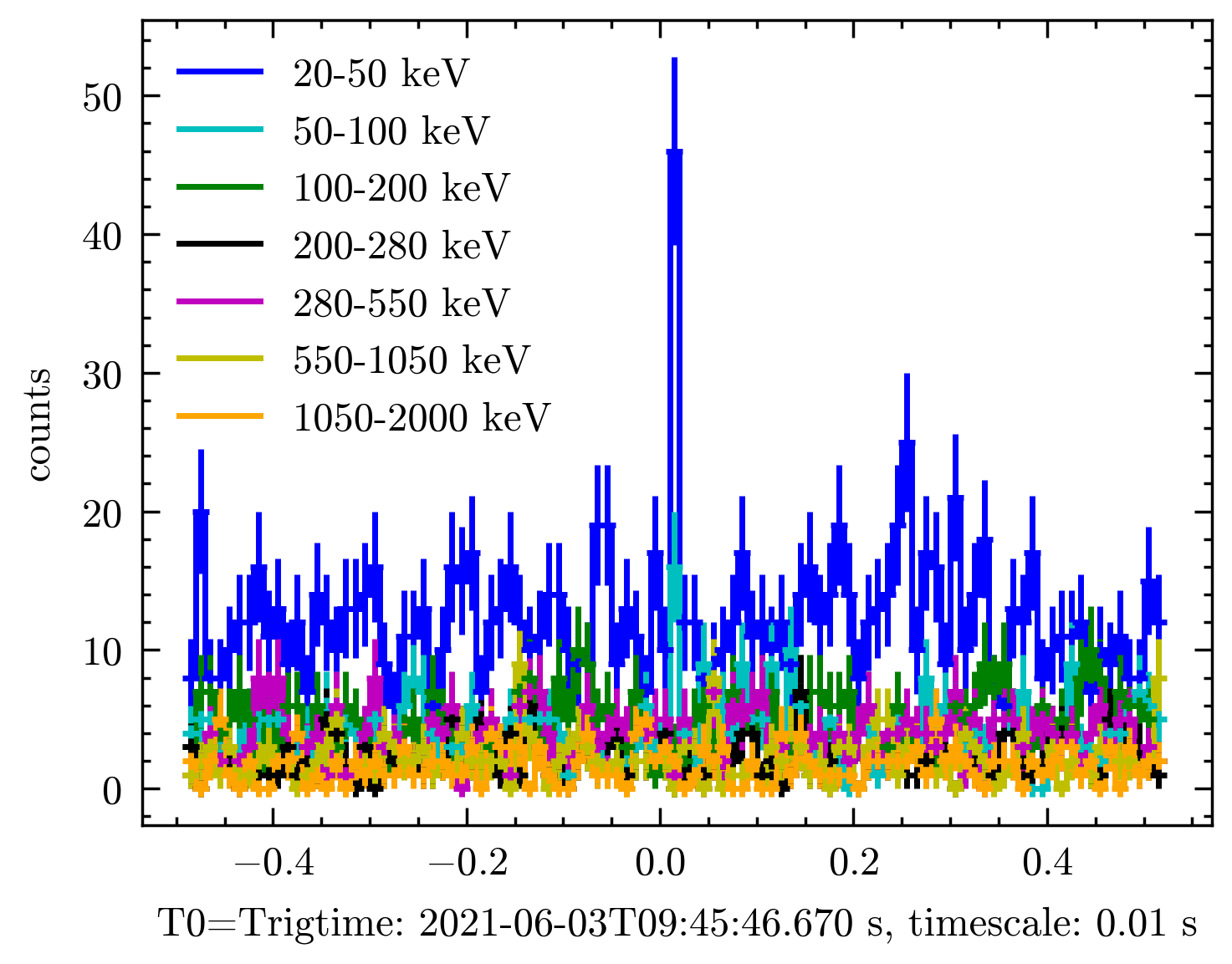} \\
		\end{tabular}
		\caption{The light curves of a SGR J1555.2-5402 burst found by targeted search with different energy ranges. Data from GECAM GRD Detector \#22, \#13, \#21, \#12, \#5.}
		\label{figure_SGR1555LightCurve}
\end{figure}

\end{multicols}

\begin{onecolumn}
\begin{ThreePartTable} 
    \begin{longtable}{cccc}
    \caption{The detailed results of ground search for GRBs} \label{Tab4} \\
    \hline
    GRB IDs & Trigger UTC & Trigger Timescale (s)  &  Observations from other GRB missions$^{1}$ \\
    \hline
    GRB 210204A	&	2021-02-04T06:29:21.950	&	2	&	GBM,	BAT,	Konus-Wind	\\
    GRB 210207B	&	2021-02-07T21:52:10.000	&	2	&	BAT,	Konus-Wind,	AGILE	\\
    GRB 210213A	&	2021-02-13T17:43:11.950	&	2	&	GBM			\\
    GRB 210228A	&	2021-02-28T06:38:31.950	&	1	&	...			\\
    GRB 210307B	&	2021-03-07T05:56:38.850	&	0.1	&	...			\\
    GRB 210307A$^{2}$	&	2021-03-07T08:42:40.000	&	2	&	BAT			\\
    GRB 210317A	&	2021-03-17T09:07:19.500	&	1	&	GBM			\\
    GRB 210328A	&	2021-03-28T20:45:18.450	&	1	&	GBM			\\
    GRB 210330A	&	2021-03-30T12:45:46.200	&	0.5	&	...			\\
    GRB 210401A$^{2}$	&	2021-04-01T23:17:08.000	&	4	&	GBM,	Konus-Wind		\\
    GRB 210409A	&	2021-04-09T21:28:07.250	&	0.1	&	...			\\
    GRB 210413A	&	2021-04-13T01:07:24.000	&	0.5	&	...			\\
    GRB 210421A$^{2}$	&	2021-04-21T00:27:15.000	&	4	&	BAT			\\
    GRB 210425A	&	2021-04-25T07:07:04.000	&	0.1	&	...			\\
    GRB 210427A	&	2021-04-27T04:57:12.950	&	0.1	&	GBM			\\
    GRB 210430B$^{2}$	&	2021-04-30T21:42:17.500	&	0.2	&	GBM			\\
    GRB 210511B	&	2021-05-11T11:26:39.800	&	0.2	&	GBM,	Konus-Wind,	AGILE	\\
    GRB 210516A	&	2021-05-16T23:34:46.400	&	0.05	&	GBM			\\
    GRB 210520A	&	2021-05-20T19:07:02.750	&	0.2	&	GBM,	BAT		\\
    GRB 210522A	&	2021-05-22T22:11:44.450	&	0.1	&	...			\\
    GRB 210529A$^{2}$	&	2021-05-29T16:19:24.800	&	0.2	&	GBM			\\
    GRB 210602B	&	2021-06-02T20:46:02.950	&	2	&	...			\\
    GRB 210606B	&	2021-06-06T22:40:39.000	&	2	&	GBM,	Konus-Wind,	AGILE	\\
    GRB 210619B	&	2021-06-19T23:59:25.550	&	0.1	&	GBM,	BAT,	Konus-Wind	\\
    GRB 210622B	&	2021-06-22T10:33:02.750	&	0.5	&	GBM			\\
    GRB 210627B	&	2021-06-27T17:57:21.550	&	0.5	&	GBM			\\
    GRB 210719A	&	2021-07-19T02:24:57.000	&	2	&	...			\\
    GRB 210822A	&	2021-08-22T09:18:18.000	&	0.05	&	BAT			\\
    GRB 210827B	&	2021-08-27T10:10:16.250	&	0.5	&	GBM			\\
    GRB 210909B	&	2021-09-09T20:02:36.700	&	0.5	&	...			\\
    GRB 210912C$^{2}$	&	2021-09-12T16:52:07.950	&	0.1	&	...			\\
    GRB 210923B$^{2}$	&	2021-09-23T02:11:29.150	&	0.05	&	...			\\
    GRB 210926A	&	2021-09-26T20:52:28.300	&	1	&	...			\\
    GRB 210927B	&	2021-09-27T23:54:42.000	&	2	&	...			\\
    GRB 211022A	&	2021-10-22T00:47:28.950	&	1	&	...			\\
    GRB 211024A$^{2}$	&	2021-10-24T01:34:11.400	&	0.05	&	GBM,	BAT		\\
    GRB 211102B	&	2021-11-02T14:05:35.200	&	0.5	&	GBM			\\
    GRB 211105A	&	2021-11-05T04:34:32.050	&	0.2	&	BAT,	AGILE		\\
    GRB 211106A$^{2}$	&	2021-11-06T04:37:31.200	&	0.05	&	GBM,	BAT,	Konus-Wind	\\
    GRB 211109C	&	2021-11-09T07:51:01.550	&	0.5	&	...			\\
    GRB 211110A	&	2021-11-10T03:26:28.450	&	0.5	&	...			\\
    GRB 211120A	&	2021-11-20T23:05:20.450	&	0.1	&	Konus-Wind,	AGILE		\\
    GRB 211124A$^{2}$	&	2021-11-24T02:29:53.900	&	0.2	&	GBM,	BAT		\\
    GRB 211201A$^{2}$	&	2021-12-01T20:43:05.950	&	4	&	GBM,	BAT		\\
    GRB 211204C	&	2021-12-04T21:36:33.000	&	2	&	GBM,	BAT		\\
    GRB 211207A$^{2}$	&	2021-12-07T20:53:00.950	&	4	&	GBM,	BAT		\\
    GRB 211211B	&	2021-12-11T21:48:42.000	&	2	&	GBM			\\
    GRB 211216A	&	2021-12-16T06:45:55.750	&	0.5	&	GBM,	BAT		\\
    GRB 211216B	&	2021-12-16T13:21:07.950	&	2	&	GBM			\\
    GRB 211217A	&	2021-12-17T07:04:29.600	&	0.1	&	...			\\
    GRB 211223A	&	2021-12-23T02:41:18.850	&	0.2	&	...			\\
    GRB 211229A	&	2021-12-29T03:29:29.250	&	0.5	&	GBM,	BAT		\\
    GRB 211229B	&	2021-12-29T22:18:28.000	&	4	&	GBM,	BAT,	AGILE	\\
    GRB 211231A	&	2021-12-31T07:00:34.800	&	0.2	&	GBM			\\
    \hline
\end{longtable}
\begin{tablenotes}
    \footnotesize
    \item[1] Observations from  other GRB missions reported in GCN circular.
    \item[2] The GRBs did not trigger the detection threshold in the on-board search.
\end{tablenotes}
\end{ThreePartTable}
\end{onecolumn}

\begin{multicols}{2}

Terrestrial-Gamma Flashes (TGFs) are short intense gamma-ray flashes produced in the lightning process in the atmosphere of the Earth \cite{1994STIN...9611316F,2017Natur.551..481E}. The duration of TGFs ranges from sub-millisecond to several milliseconds. The spectra of TGFs usually extend from hundreds of keV up to tens of MeV \cite{1997JGR...102.9659N,2013JGRA..118.3805B}, much harder than GRBs (tens of keV to several MeV) and SGR bursts (several of keV to hundreds of keV). Zhao et al. \cite{2022esoar.54151023Z} reported a dedicated burst search algorithm for TGFs, which assumes the background following the Poisson distribution and then calculates the probability of the observed counts from background fluctuation. This search is implemented for the following time-scales: 50 $\mu$s, 100 $\mu$s, 250 $\mu$s, 500 $\mu$s, 1 ms, 2 ms and 4 ms.

On March 29, 2021, the automatic detection of a TGF was found by GECAM ground search, which demonstrates that ground search also have the ability to find relatively bright TGFs.
Figure \ref{figure_TGFLightCurve1} shows the light curves of this detected TGF, illustrating that a statistically significant rate increase above the background rate was detected in the energy of about hundreds of keV to several MeV within 50 ms.

\subsection{Targeted Search} \label{target_search}

The targeted search was developed to examine GECAM data for important gamma-ray transients which did not triggered GECAM on-board search and ground blind search, but were detected by other instruments (e.g., \textit{Insight}-HXMT, \textit{Fermi}/GBM, \textit{Swift}/BAT, LIGO, Virgo, and KAGRA). The targeted search time window can be set by BA (e.g.,30 s ), which is centered on the burst trigger time from other instruments. BA can use the default values of other search parameters, including data type (BTIME or EVT data), bin-widths, location, spectra of bursts, search threshold. These values can also be changed depending on the types of gamma-ray transients.

SGR J1555.2-5402 is a new magnetar in our galaxy, which first triggered \textit{Swift}/BAT through a short hard X-ray burst on June 3rd, 2021 \cite{2021GCN.30120....1P}. This X-ray burst did not trigger either GECAM on-board search or ground blind search. Using the precise location provided by BAT, we find that it is visible to GECAM. The targeted search follow-up of this burst used the same threshold as the blind search (see Table \ref{Tab1}). Considering the duration of the magnetar burst, the minimum search timescale is set to 10 ms. With high time resolution event data, the search was performed on binned data with bin-widths ranging from 10 ms to 4 s. With this procedure, there is a trigger found by this targeted pipeline. Both the trigger time of 2021-06-03T09:45:46.670 UTC (see Figure \ref{figure_SGR1555LightCurve}) and location are consistent with those of \textit{Swift}/BAT. The results are also presented in \cite{2021GCN.30140....1C}.

\section{DISCUSSION AND CONCLUSION}
\label{sect:discussion}

In this paper, we present a dedicated GECAM ground burst search system (see Figure \ref{figure_GroundSearchFlow}) including automatic mode and manual mode to search GECAM data in depth for gamma-ray transients (e.g, GRBs, SGRs, gamma-ray counterparts of GW, FRBs and neutrinos). Both blind search and targeted search using a coherent method have been implemented in this pipeline. For blind search, the fully automatic data analysis pipeline once enabled, allows the immediate use of this method for transient detections in the continuous data stream from the GECAM detector. Gamma-ray transients can be found using relatively high thresholds (Table \ref{Tab1}) to avoid excessive false triggers. As for targeted search, sub-threshold counterparts to GW, FRBs and neutrinos can be found with manual defined search threshold. With this pipeline, quick-look data is processed for BA to classify triggers and eliminate false triggers caused by instrumental effects or background fluctuations.

The ground search using the weighted light curve based on the coherent method can substantially enhance the detection significance (see Cai C. et al. for more details \cite{10.1093/mnras/stab2760}). The sensitivity of GECAM on-board search and ground search is estimated using original light curves and weighted light curves, respectively. Assuming a burst with soft Band spectrum \cite{Band:1993}, a typical duration of 20 s, and location of theta = 3 deg and phi = 45 deg (in the GECAM payload coordinate system), we can calculate the expected counts.
Using the observed background and expected counts, we can estimate the corresponding normalization of spetrum and flux when SNR equals 3-sigma. With the original light curves, the detection sensitivity of GECAM is about 7.2E-08 erg cm$^{-2}$ s$^{-1}$ (10 keV $-$ 1000 keV), while with the weighted light
curves, the detection sensitivity is increased to about 1.1E-08 erg cm$^{-2}$ s$^{-1}$ (10 keV $-$ 1000 keV).

The results outlined in section \ref{sect:Performance} demonstrate that the ground search is an effective and reliable pipeline to unveil much weaker bursts which did not trigger the on-board search. For blind search, we find a total of 54 GRBs, 63 SGRs, 5 XRB candidates, 8 TGFs, more than 100 SFs, Earth occultation of known sources and particle events, and dozens of instrument effect triggers (see Figure \ref{figure_TriggerStatistics} and Table \ref{Tab2} for details). 
We take light curves as examples to show the characters of each type of triggers (see Figure \ref{figure_4U 0614+09} $-$ Figure \ref{figure_TGFLightCurve1}). 
We check GECAM GRD light curves of GRBs which did not trigger the on-board search and find that these bursts with less signals of each GRD can be unveiled by ground search through coherently accumulating GECAM multiple-detector multiple-channel data.

The targeted search using lower threshold and/or suitable search timescale can also successfully recover some interesting and import bursts (e.g., magnetar bursts, Figure\ref{figure_SGR1555LightCurve}), that are too weak to trigger the blind search.

The ground search system maximizes the potential of GECAM observations in the era of multi-messenger and multi-wavelength astronomy. Therefore, it is critically important to search GECAM data for high-energy transients associated with GW events, FRBs, or neutrinos, and we will report these results in the next paper.

\Acknowledgements{This work is supported by the National Natural Science Foundation of China 
(Grant No. 12303045, 12273042, 12373047) 
, the National Key R\&D Program of China (
2022YFF0711404), 
the Strategic Priority Research Program of Chinese Academy of Sciences (Grant No. XDA15360102, XDA15360300, XDA15052700). 
The GECAM (Huairou-1) mission is supported by the Strategic Priority Research Program on Space Science (Grant No. XDA15360000) of the Chinese Academy of Sciences.
C.C. acknowledges the support from the Natural Science Foundation of Hebei Province (No. A2023205020) and the Science Foundation of Hebei Normal University (No. L2023B11).} 

\InterestConflict{The authors declare that they have no conflict of interest.}



\normalem

\end{multicols}

\end{document}